\documentclass[aps,pra,reprint,superscriptaddress]{revtex4-2}

\usepackage{dcolumn}
\usepackage{bm}
\usepackage{hyperref}
\usepackage{verbatim}

\usepackage{graphicx} 
\usepackage{graphicx}
\usepackage{amssymb}
\usepackage{amsmath}
\usepackage{epsf}
\usepackage{caption, subcaption, floatrow}

\usepackage[dvipsnames]{xcolor}

\usepackage[figcolor=white]{todonotes}

\date{}
\begin{document}

\title{Transparent Boundary Conditions for the Heat Equation on Metric Graphs}

\author{Jasur Matrasulov}
\affiliation{
Department of Electronics and Radio Engineering, 
Tashkent University of Information Technologies,
Amir Temur Str.~108,
Tashkent 100084, Uzbekistan
}

\author{Jambul Yusupov}
\affiliation{
Kimyo International University in Tashkent,
Shota Rustaveli Str.~156,
Tashkent 100121, Uzbekistan
}

\author{Matthias Ehrhardt}
\affiliation{
Applied and Computational Mathematics,
University of Wuppertal,
Gau{\ss}strasse~20,
42119 Wuppertal, Germany
}


\begin{abstract}
We study transparent boundary conditions (TBCs) for the time-depen\-dent heat equation in a branching, quasi-one-dimensional domain modeled as a metric star graph. 
By combining the classical concept of TBCs with the theory of partial differential equations (PDEs) on networks, we derive an exact vertex condition that ensures unobstructed thermal flow across junctions. 
Specifically, we derive a sum rule for the diffusion coefficients 
that eliminates thermal backflow at the vertex. 
We demonstrate the validity of this analytical model numerically using a Crank-Nicolson finite-difference method, which confirms smooth, unobstructed heat propagation through the network. 
These results provide a practical mathematical framework for tunable control and optimization of thermal diffusion in low-dimensional structures.
 In particular, combining the well-known concept of the TBCs
 and theory for PDE on metric graphs,  we propose a mathematical model providing a control tool for thermal diffusion in networks. 
\end{abstract}
\maketitle

\section{Introduction}

Tunable heat transport in low-dimensional structures is of great importance for many areas of modern physics and emerging technologies. Applications spanning nano\-technology, advanced electronics, thermoelectric energy harvesting, and microfluidics require precise control over thermal dissipation, containment, and propagation. 
In nanoelectronics, for instance, the rapid scaling of electronic components has intensified the challenge of mana\-ging highly localized thermal hotspots. 
In nanostructured carbon materials, such as graphene-based transistors, carbon nanotubes, and complex device interconnects, efficient heat removal is essential for maintaining device integrity, ensuring operational stability, and preventing severe thermal degradation. 
As structural dimensions shrink below the phonon mean free path, traditional Fourier law descriptions break down, necessitating sophisticated mathematical and physical frameworks to model non-Fourier ballistic-diffusive transport along these low-dimensional pathways.  

The optimization of thermoelectric energy conversion platforms relies 
on tailoring thermal behaviors at the nanoscale.
The efficiency of thermoelectric generation is governed by the dimensionless figure of merit, which requires a delicate decoupling of electronic and thermal transport pathways \cite{balandin, cahill, chouly, Dhar2008, Lepri2016}. 
To maximize this efficiency, 
one must suppress lattice thermal conductivity 
via structured material interfaces, localized boundary scattering, and potential barriers without simultaneously hindering electrical conductivity. Low-dimensional branched architectures, quantum wire networks, and superlattices offer highly tunable environments where phone engineering can minimize unwanted heat propagation \cite{Matrasulov2023}.  

Beyond solid-state devices, local temperature control plays a foundational role in microfluidics and lab-on-a-chip platforms. 
In these systems, precise thermal mana\-gement within microscopic networks and channels is indispensable for stabilizing rapid chemical reactions, enabling highly accurate nucleic acid amplification, and driving thermophoresis, the directed motion of particles induced by thermal gradients. 
Effective heat removal and routing are 
critical for managing the thermal output generated by active micro-pumping elements and integrated micro-sensors embedded in localized polymer networks.  

A key engineering challenge across these diverse fields is to minimize thermal energy losses during directional heat transfer and to completely eliminate unwanted backflow or reflections at structural interfaces, junctions, and material boundaries. 
Over the past few decades, partial differential equations (PDEs) defined on quasi-one-dimensional branched domains, commonly referred to as metric graphs or networks, have emerged as a powerful mathematical tool for modeling wave propagation and diffusion processes  \cite{Kottos1999, Hul2004,Kuchment2004, Gnutzmann2006, ExnerKovarik2015, Exner1988, Kostrykin1999, Noja2015, Adami2016, Kairzhan2018,  Babajanov2018,YusupovChemPhys2020, MatrasulovEPL2020, Sabirov2020}. 
While wave dynamics, quantum waveguides (such as the Schr\"odinger equation), and relativistic particle transport on metric graphs have been extensively analyzed over recent decades \cite{yusupov2020, yusupov2019}, modeling and controlling thermal diffusion on these interconnected network topologies remains a highly active and challenging frontier. 

When solving the heat equation numerically on unbounded, semi-infinite, or exceedingly large domains, a major computational bottleneck is limiting the spatial domain without introducing non-physical, artificial boundary reflections. 
For standard 
1D domains, transparent boundary conditions (TBCs) 
effectively resolve this issue by allowing thermal profiles to exit the computational window seamlessly as if diffusing into an infinite environment \cite{antoine}.
However, extending this well-established TBC framework to complex, interconnected networks presents unique mathematical challenges, particularly regarding how the thermal flux splits at a graph vertex.  

In this paper, we address this issue by combining the TBC approach with PDE analysis on metric graphs. 
To do so, we formulate a mathematical model that describes the behavior of thermal diffusion across a junction. Importantly, we derive an explicit analytical constraint on the network's material properties, specifically expressed as an exact sum rule for the diffusion coefficients, to ensure completely unobstructed heat transmission from an incoming channel to outgoing branches.  

This work is organized as follows: 
Section~\ref{sec2} briefly reviews the classical formulation of TBCs for the time-dependent heat equation on a real line using the Laplace transform. 
Section~\ref{sec3} extends this approach to a star graph topology and derives the continuity and Kirchhoff-type vertex conditions necessary for transparent transmission. 
Section~\ref{sec4} details the numerical implementation of our model using a Crank-Nicolson (CN) finite-difference method (FDM) and visualizes the resulting reflectionless heat propagation. 
Finally, Section~\ref{sec5} offers a brief summary and concluding remarks. 

\section{TBC for the Heat Equation on a Line}\label{sec2}





The one-dimensional heat equation governs the temporal evolution of a temperature profile $u(x, t)$. 
When simulating this process on a truncated or bounded spatial domain, an ideal TBC ensures that the temperature profile does not ``notice" the artificial boundary. 
In other words, the solution restricted to the computational domain must be identical to the solution obtained if the domain extended across the entire real line $\mathbb{R}$.  

To illustrate the classical derivation \cite{ehrhardt, ehrhardt96, HanHuang2002a}, consider the heat equation on the real line:
\begin{equation} \label{eq:heat}
    \frac{\partial u}{\partial t}
=\kappa^2\, \frac{\partial^2 u}{\partial x^2},
\quad -\infty<x<\infty,\; t>0,
\end{equation}
where $\kappa^2$ denotes the thermal diffusivity constant. 
For a Gaussian initial temperature profile of the form
\begin{equation}
u(x,0)=u_0
\exp\left[-\frac{(x-x_0)^2}{2\sigma^2}\right],
\label{initial_gaussian}
\end{equation}
where $u_0$ is the initial peak temperature, $x_0$ specifies the center of the profile, and $\sigma$ characterizes its width, the solution to Eq.~\eqref{eq:heat} can be written as
\begin{equation}
u(x,t)=
\frac{1}
{\sqrt{2\pi\left(\sigma^2+2\kappa^2 t\right)}}
\exp\left[
-\frac{(x-x_0)^2}
{2\left(\sigma^2+2\kappa^2 t\right)}
\right].
\end{equation}
where $C_1$ and $C_2$ are constants determined by the physical boundary conditions.  
The problem of constructing exact, approximate, and local transparent
or artificial boundary conditions for the heat equation and related
parabolic problems on unbounded domains has been widely studied
\cite{HanHuang2002b, WuSun2004, Zheng2007, DitkowskiSuhov2008, 
WuZhang2011, SuhovDitkowski2011, zisowsky}.
Following the classical approach established in \cite{ehrhardt, ehrhardt96, HanHuang2002a}, we outline the derivation of a TBC at an artificial boundary located at $x=0$ for a semi-infinite ('interior') computational domain $x<0$.

Our objective is to derive a relation between the Diri\-chlet data $u(0,t)$ and the Neumann data (thermal flux) $\frac{\partial u}{\partial x}\big|_{x=0}$. 
Assuming a quiescent initial state $u(x,0)=0$, for $x>0$, we apply the 
Laplace transform with respect to time
\begin{equation*}
  \hat{u}(x,s)=\mathcal{L}{u(x,t)}
  =\int_0^\infty u(x,t) \,\mathrm{e}^{-st}\,dt, 
\end{equation*}
which transforms the PDE~\eqref{eq:heat} into a second-order 
ordinary differential equation (ODE) in the spatial variable defined on the \textit{exterior domain} $x>0$:
\begin{equation}\label{eq:ODE}
    \frac{d^2 \hat{u}}{dx^2} - \frac{s}{\kappa^2}\,\hat{u} = 0,\quad x>0.
\end{equation}
Choosing the physical (bounded) solution as $x\to\infty$ yields:
\begin{equation}\label{eq:4}
    \hat{u}(x,s) = \hat{u}(0,s)\,\mathrm{e}^{-\frac{\sqrt{s}}{\kappa}x}.
\end{equation}
Differentiating \eqref{eq:4} with respect to $x$ and evaluating the result at 
$x=0$ provides the explicit flux relationship:
\begin{equation}\label{eq:flux}
    \frac{d\hat{u}}{dx}(0,s) = -\frac{\sqrt{s}}{\kappa} \,\hat{u}(0,s).
\end{equation}
Applying the inverse Laplace transform to \eqref{eq:flux},
where the multiplier $\sqrt{s}$ maps to a Riemann-Liouville fractional derivative of order $1/2$ in the time domain,
we obtain the 
TBC for a thermal profile on the exterior domain $x\ge0$:
\begin{equation}\label{eq:TBC}
    \frac{\partial u}{\partial x}\bigg|_{x=0} 
    = -\frac{1}{\kappa\sqrt{\pi}}\int_{0}^{t}\frac{1}{\sqrt{t-\lambda}}\frac{\partial u(0,\lambda)}{\partial \lambda}\,d\lambda.
\end{equation}
We note that conversely, if we consider a 
thermal profile 
diffusing inside or outside
the domain $x\le0$, the sign of the spatial derivative is reversed due to the outward unit normal vector pointing in the positive $x$-direction:
\begin{equation}\label{eq:classicalTBC}
    \frac{\partial u}{\partial x}\bigg|_{x=0} = +\frac{1}{\kappa\sqrt{\pi}}\int_{0}^{t}\frac{1}{\sqrt{t-\lambda}}\frac{\partial u(0,\lambda)}{\partial \lambda}\,d\lambda.
\end{equation}
This non-local in time, mildly singular, convolution-type boundary operator \eqref{eq:TBC} serves as the foundational building block for analyzing network junctions,
which we extend to a star graph topology in the next section.

\section{Extension to a Star Graph}\label{sec3}
%
The design of TBCs for evolution equations on metric graphs implies their imposing at the graph vertices (nodes), as so-called 'branching conditions'. 
The latter allows to consider graphs or networks with ``transparent” vertices, or ``transparent graphs”. 
From the viewpoint of (signal, energy, heat, etc.) transport, such a transparency implies minimization of the losses during the transfer. 
Within such an approach, for the heat equation on graphs, the TBC concept can be an effective tool for controlling of thermal diffusion in networks. 
Below we demonstrate this approach for simplest graph topo\-logy, star graph, although extensions of this approach for other (arbitrary) topology is straight forward.

Consider a simple metric star graph consisting of three semi-infinite bonds $b_j$, $j=1,2,3$, where 
$b_1 \sim (-\infty,0]$, and $b_{2,3} \sim [0,+\infty)$.
On each individual bond $b_j$, the temporal evolution of the temperature profile $u_j(x,t)$ is governed by the localized heat equation:
\begin{equation}\label{eq:heat2}
  \frac{\partial u_j}{\partial t} = \kappa_j^2\,\frac{\partial^2 u_j}{\partial x^2}, \quad x\in b_j,\; t>0,
\end{equation}
where $\kappa_j^2$ represents the thermal diffusivity of bond $b_j$.
At the central junction ($x=0$),
the sub-problems are coupled using the standard continuity condition
\begin{equation}\label{eq:continuity}
    u_1(0,t)=u_2(0,t)=u_3(0,t),
\end{equation}
together with the \textit{Kirchhoff-type flux balance rule}
\begin{equation}\label{eq:kirchhoff}
  \kappa_1^2\,\frac{\partial u_1}{\partial x}\Big|_{x=0}
  =\kappa_2^2\,\frac{\partial u_2}{\partial x}\Big|_{x=0}
   +\kappa_3^2\,\frac{\partial u_3}{\partial x}\Big|_{x=0}.
\end{equation}

To analyze the diffusion process across the junction, 
we decouple the network into an ``interior" problem on the incoming branch $b_1$ and ``exterior" problems on the outgoing branches $b_{2,3}$. 
The interior problem on $b_1 \sim (-\infty,0]$ is formulated as
\begin{equation}
   \frac{\partial u_1}{\partial t}=\kappa_1^2\,\frac{\partial^2 u_1}{\partial x^2},\quad x<0,
\end{equation}
subject to a non-zero initial thermal distribution
\begin{equation}
  u_1(x,0)=f_1^I(x),\quad x<0,
\end{equation}
and a Dirichlet-to-Neumann (DtN) operator $T$ acting at the vertex $x=0$:
\begin{equation}
   (Tu_1)(0,t)=\frac{\partial u_1}{\partial x}(0,t).
\end{equation}

Correspondingly, the ``exterior'' problems on the outgoing semi-infinite bonds $b_{2,3} \sim [0,+\infty)$
take the form
\begin{equation}
  \frac{\partial u_{2,3}}{\partial t}=\kappa_{2,3}^2\,\frac{\partial^2 u_{2,3}}{\partial x^2},\quad x>0,
\end{equation}
under a quiescent initial condition for both bonds
\begin{equation}
   u_{2,3}(x,0)=0,\quad x>0,
\end{equation}
with the corresponding vertex boundary values at $x=0$:
\begin{equation}\label{eq16}
   u_{2,3}(0,t)=g_{2,3}(t),\qquad
  \frac{\partial u_{2,3}}{\partial x}(0,t)=(Tu_{2,3})(0,t).
\end{equation}

To solve the exterior problem analytically, we apply the
Laplace transform
\begin{equation}
   \tilde{u}_{2,3}(x,s) = \int_0^\infty u_{2,3}(x,t)\,\mathrm{e}^{-st}\,dt,
\end{equation}
which yields the second-order ODEs on the bonds $b_2$, $b_3$:
\begin{equation}\label{eq:trans_ext_problem}
    s\,\tilde{u}_{2,3}(x,s)-\kappa_{2,3}^2\,\frac{\partial^2 \tilde{u}_{2,3}}{\partial x^2}(x,s)=0.
\end{equation}

The general solution to \eqref{eq:trans_ext_problem} can be written as a linear combination of exponentials (the fundamental solutions):
\begin{equation}
  \tilde{u}_{2,3}(x,s)
  = A_{2,3}\mathrm{e}^{\sqrt{\frac{s}{\kappa_{2,3}^2}}\,x}
    +B_{2,3}\mathrm{e}^{-\sqrt{\frac{s}{\kappa_{2,3}^2}}\,x}.
\end{equation}
Imposing the physical restriction that the exterior solutions must remain bounded, i.e.\ $\tilde{u}_{2,3}\in L^2(0,+\infty)$, 
requires the coefficients of the growing exponentials to vanish ($A_{2,3}=0$). 
This reduces the solution on the outgoing branches to:
\begin{equation}\label{eq21}
  \tilde{u}_{2,3}(x,s)=B_{2,3}\,\mathrm{e}^{-\sqrt{\frac{s}{\kappa_{2,3}^2}}\,x}.
\end{equation}
Evaluating \eqref{eq21} at the interface $x=0$ using \eqref{eq16} matches the integration constants to the boundary data
\begin{equation}\label{eq22}
    B_{2,3}=\tilde{u}_{2,3}(0,s)=\tilde{g}_{2,3}(s).
\end{equation}
From Eqs.~\eqref{eq21} and \eqref{eq22} we obtain
\begin{equation}\label{eq23}
   \tilde{u}_{2,3}(x,s)=\tilde{g}_{2,3}(s)\,\mathrm{e}^{-\sqrt{\frac{s}{\kappa_{2,3}^2}}\,x},
\end{equation}
and differentiating \eqref{eq23} with respect to $x$ provides the formal flux relations in the Laplace domain
\begin{equation}\label{eq24}
    \begin{split}
\frac{\partial \tilde{u}_{2,3}(x,s)}{\partial x}
&=-\sqrt{\frac{s}{\kappa_{2,3}^2}}\,\tilde{g}_{2,3}(s)
\,\mathrm{e}^{-\sqrt{\frac{s}{\kappa_{2,3}^2}}\,x}\\
&=-\sqrt{\frac{s}{\kappa_{2,3}^2}}\,\tilde{u}_{2,3}(x,s).
 \end{split}
\end{equation}
Applying the continuity condition~\eqref{eq:continuity} at $x=0$, we link the exterior solutions 
to the incoming branch state
\begin{equation}\label{eq25}
    \tilde{u}_{2,3}(0,s)=\tilde{u}_1(0,s).
\end{equation} 
Evaluating the spatial derivatives at $x=0$ via Eqs.~\eqref{eq24} and \eqref{eq25} yields
\begin{equation}\label{eq26}
  \frac{\partial \tilde{u}_{2,3}}{\partial x}(0,s)
  = -\sqrt{\frac{s}{\kappa_{2,3}^2}}\,\tilde{u}_{2,3}(0,s)
  = -\sqrt{\frac{s}{\kappa_{2,3}^2}}\,\tilde{u}_1(0,s).
\end{equation}
Next, substituting these exterior derivatives into the Kirchhoff rule~\eqref{eq:kirchhoff} and \eqref{eq26}, we find the coupled flux relation for the interior domain
\begin{equation}\label{eq27}
\begin{split}
\frac{\partial \tilde{u}_1}{\partial x}(0,s)
&=\frac{\kappa_2^2}{\kappa_1^2}\frac{\partial \tilde{u}_2}{\partial x}(0,s)
+\frac{\kappa_3^2}{\kappa_1^2}\frac{\partial \tilde{u}_3}{\partial x}(0,s)\\
&=\frac{\kappa_2+\kappa_3}{\kappa_1^2}\sqrt{s}\,\tilde{u}_1(0,s).
\end{split}
\end{equation}
Applying the inverse Laplace transform to \eqref{eq27}, we reveal the TBC for the branching interface vertex
\begin{equation}\label{eq:tbc27}
\frac{\partial u_1}{\partial x}(0,t)
=\frac{\kappa_2+\kappa_3}{\kappa_1^2\sqrt{\pi}}
\int_0^t\frac{1}{\sqrt{t-\tau}}
\frac{\partial u_1}{\partial \tau}(0,\tau)\,d\tau.
\end{equation}
By comparing \eqref{eq:tbc27} to the classical continuous TBC \eqref{eq:classicalTBC},
a structural alignment emerges. 
If the network's para\-meters satisfy the algebraic constraint
\begin{equation}\label{sum_r}
   \kappa_1 = \kappa_2 + \kappa_3,
\end{equation}
the boundary expression simplifies completely to match the unbounded continuum, ensuring perfect, unobstructed transmission across the network vertex.

\section{Numerical Results} \label{sec4}
\begin{figure}[t]
\centering
\includegraphics[width=\textwidth]{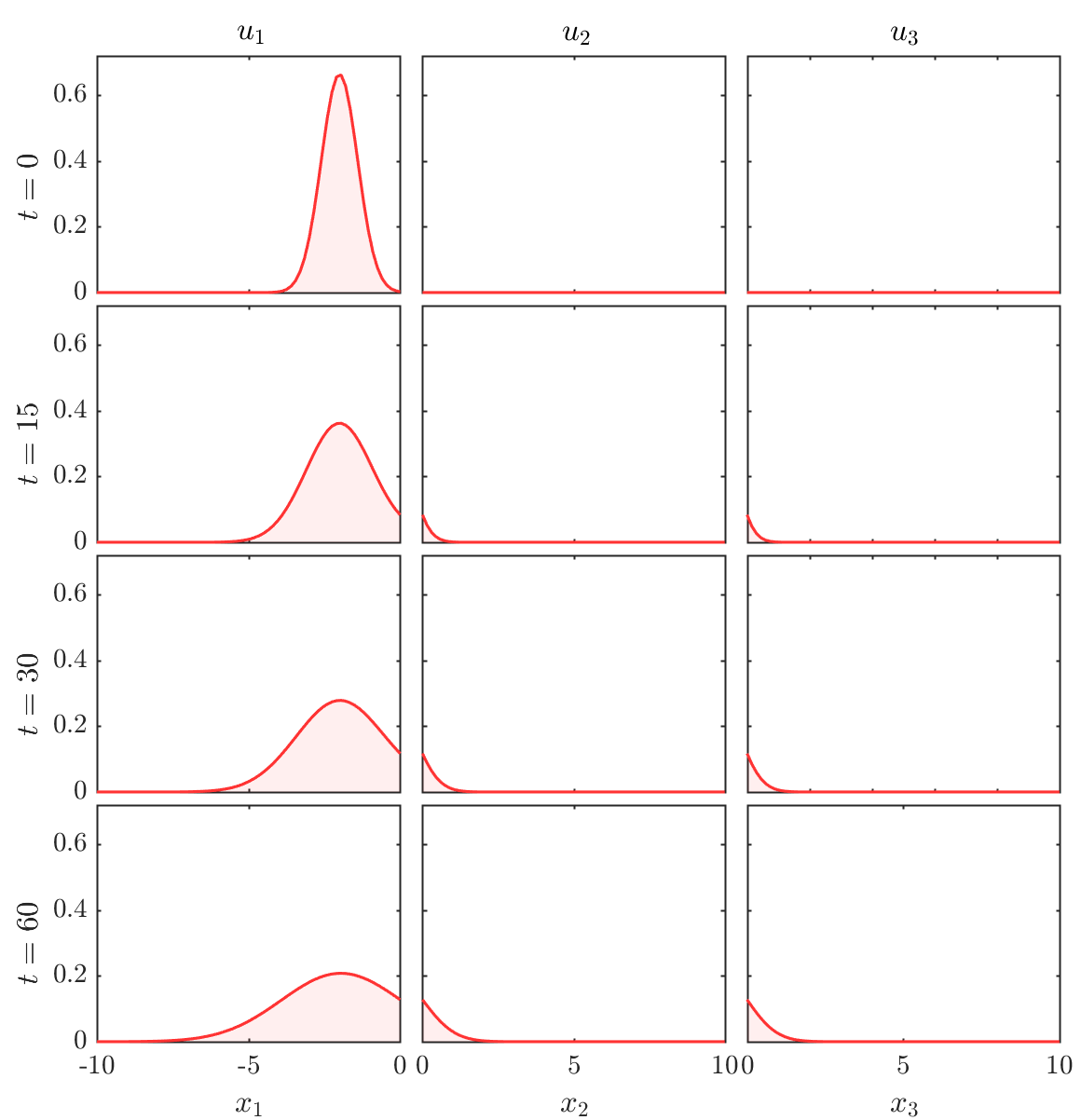}
\caption{Evolution of the temperature profile on the star graph in the reflectionless regime 
corresponding to the sum rule $\kappa_1=\kappa_2+\kappa_3=0.169$, with 
$\kappa_2=0.089$, and $\kappa_3=0.08$. 
The left, middle, and right columns show the temperature distributions on bonds 1, 2, and 3.} 
\label{fig:temp_profile}
\end{figure}

Here, we present the details and results of numerically solving the heat equation on a metric star graph. 
Unlike the standard treatment of unbounded domains, our approach does not require discretizing a nonlocal convolution-type TBC at an external boundary. 
Instead, we truncate the computational domains of the outgoing branches far enough from the vertex.
The exact transparent transmission behavior across the junction is captured by enforcing the continuity and Kirchhoff-type rules given by 
~\eqref{eq:continuity} and \eqref{eq:kirchhoff} directly at the vertex node. 
For each individual bond, $b_j$, the localized heat equation \eqref{eq:heat2} for $u_j$ is discretized using the unconditionally stable Crank-Nicolson (CN) FDM 
\begin{align}
   \frac{u_{j,i}^{n+1}-u_{j,i}^n}{\Delta t}
  =\frac{\kappa_j^2}{2(\Delta x)^2}\bigl[( & u_{j,i+1}^{n+1}-2u_{j,i}^{n+1}+u_{j,i-1}^{n+1})\label{CN}\\
   & +(u_{j,i+1}^n-2u_{j,i}^n+u_{j,i-1}^n)\bigr], \nonumber
\end{align}
with the pointwise approximation $u_{j,i}^n\approx u_j(x_i,t_n)$,
$t_n=n\Delta t$, $x_i=(i-J)\Delta x$ for bond 1 and $x_i=i\Delta x$ 
for bonds 2 and 3, where $i=0,1,\dots,J$.
By introducing the \textit{localized parabolic mesh ratio}
\begin{equation} 
   \lambda_j=\kappa_j^2\,\frac{\Delta t}{(\Delta x)^2},
\end{equation}
the CN scheme \eqref{CN} can be rewritten as a tridiagonal system 
for the interior spatial nodes of each bond:
\begin{multline}\label{CN_matrix}
  -\frac{\lambda_j}{2}\,u_{j,i-1}^{n+1}
  +(1+\lambda_j)\,u_{j,i}^{n+1}-\frac{\lambda_j}{2}\,u_{j,i+1}^{n+1}\\
   =\frac{\lambda_j}{2}\,u_{j,i-1}^n+(1-\lambda_j)\,u_{j,i}^n+\frac{\lambda_j}{2}\,u_{i+1}^n.   
\end{multline}

\begin{figure}[t]
\centering
\includegraphics[width=\textwidth]{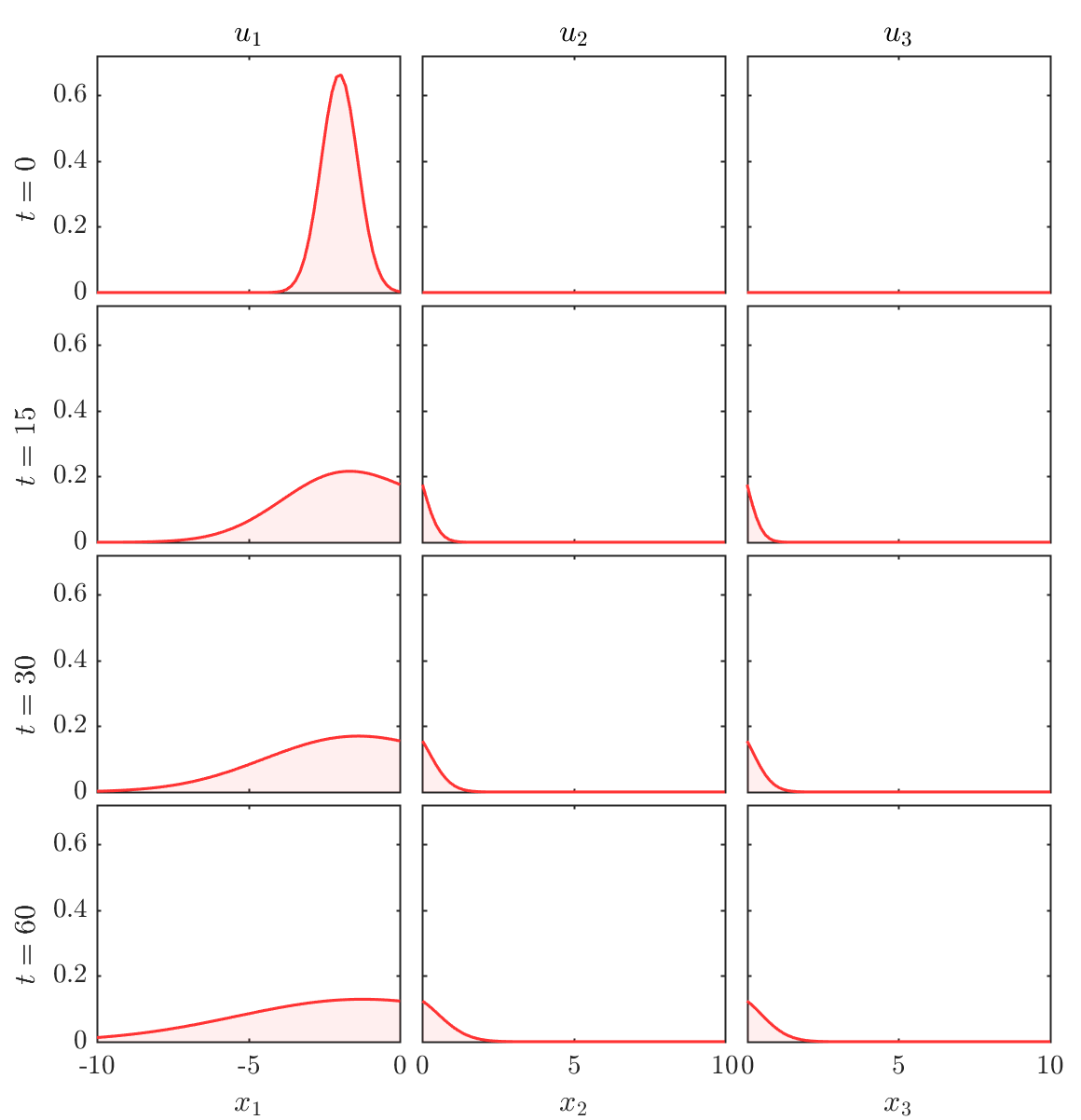}
\caption{Evolution of the temperature profile on the star graph when the sum rule~\eqref{sum_r} is broken, i.e.\ when 
$\kappa_1\neq\kappa_2+\kappa_3$, with 
$\kappa_1=0.35$, $\kappa_2=0.089$, and $\kappa_3=0.08$. 
The left, middle, and right columns show the temperature distributions on bonds 1, 2, and 3.} 
\label{fig:temp_profile2}
\end{figure}

At the central junction, the Crank-Nicolson schemes on the three bonds
are coupled through the continuity and Kirchhoff conditions. Since the
vertex corresponds to $x_J=0$ on bond 1 and $x_1=0$ on bonds 2 and 3, the spatial derivatives in the Kirchhoff condition are approximated by
one-sided finite differences as
\begin{equation}\label{discrete_kirchhoff}
\kappa_1^2\frac{u_{1,J}^{n+1}-u_{1,J-1}^{n+1}}{\Delta x}
-\kappa_2^2\frac{u_{2,2}^{n+1}-u_{2,1}^{n+1}}{\Delta x}
-\kappa_3^2\frac{u_{3,2}^{n+1}-u_{3,1}^{n+1}}{\Delta x}=0.
\end{equation}

The continuity condition is imposed as
\begin{equation}\label{discrete_continuity}
    u_{1,J}^{n+1}=u_{2,1}^{n+1}=u_{3,1}^{n+1}.
\end{equation}
These conditions replace the finite-difference equations at the vertex nodes, coupling the three bondwise Crank-Nicolson systems into a single sparse linear system at each time step. The resulting system is solved using an LU factorization of the coefficient matrix.

To validate the analytical reflectionless condition, we selected a spatial step size of $\Delta x=0.05$ and a time step of $\Delta t=0.005$. 
The diffusion parameters are taken as $\kappa_1=0.169$, $\kappa_2=0.089$, and $\kappa_3=0.08$, so that the sum rule $\kappa_1=\kappa_2+\kappa_3$ is satisfied. 
The initial temperature profile is chosen as a Gaussian pulse entirely localized on the incoming bond $b_1$, while the outgoing bonds $b_2$ and $b_3$ are initially at a quiescent state:
\begin{equation*}
    u_1(x,0) = \mathrm{e}^{-(x+2)^2/0.2}, \quad 
    u_2(x,0) = 0, \quad u_3(x,0) = 0.
\end{equation*}

\begin{figure}[t]
\centering
\includegraphics[width=1.\textwidth]{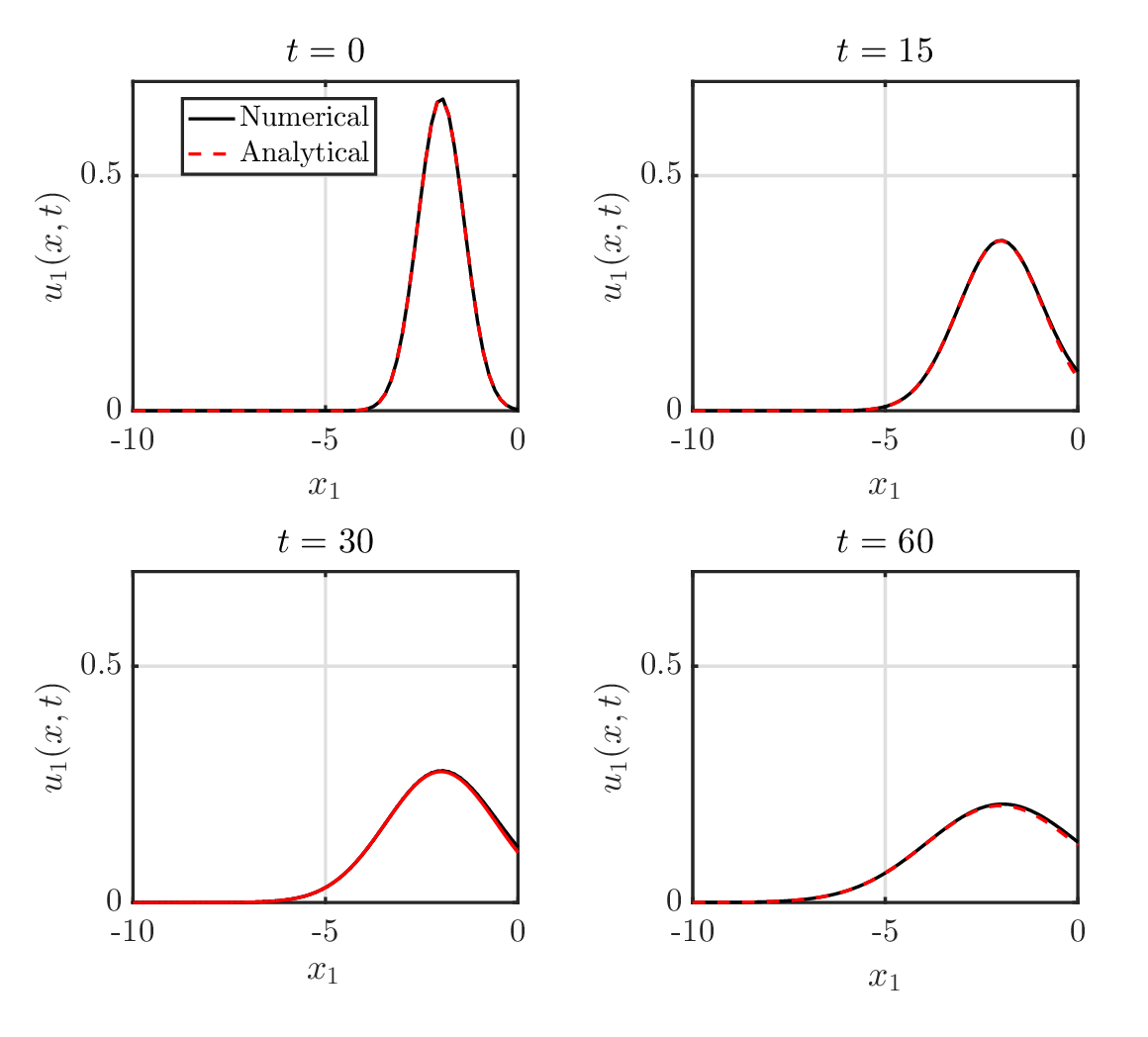}
\caption{Comparison of the numerical and analytical temperature profiles on bond~1,
$x_1\in[-L,0]$, at $t=0$, $15$, $30$, and $60$, when the sum rule
$\kappa_1=\kappa_2+\kappa_3=0,169$ is satisfied, with 
$\kappa_2=0.089$, and $\kappa_3=0.08$. 
The solid black curves represent the numerical CN
solution, and the dashed red curves show the analytical
Gaussian solution on the real line.}
\label{fig:tmprs}
\end{figure}

\begin{figure}[t]
\centering
\includegraphics[width=1.\textwidth]{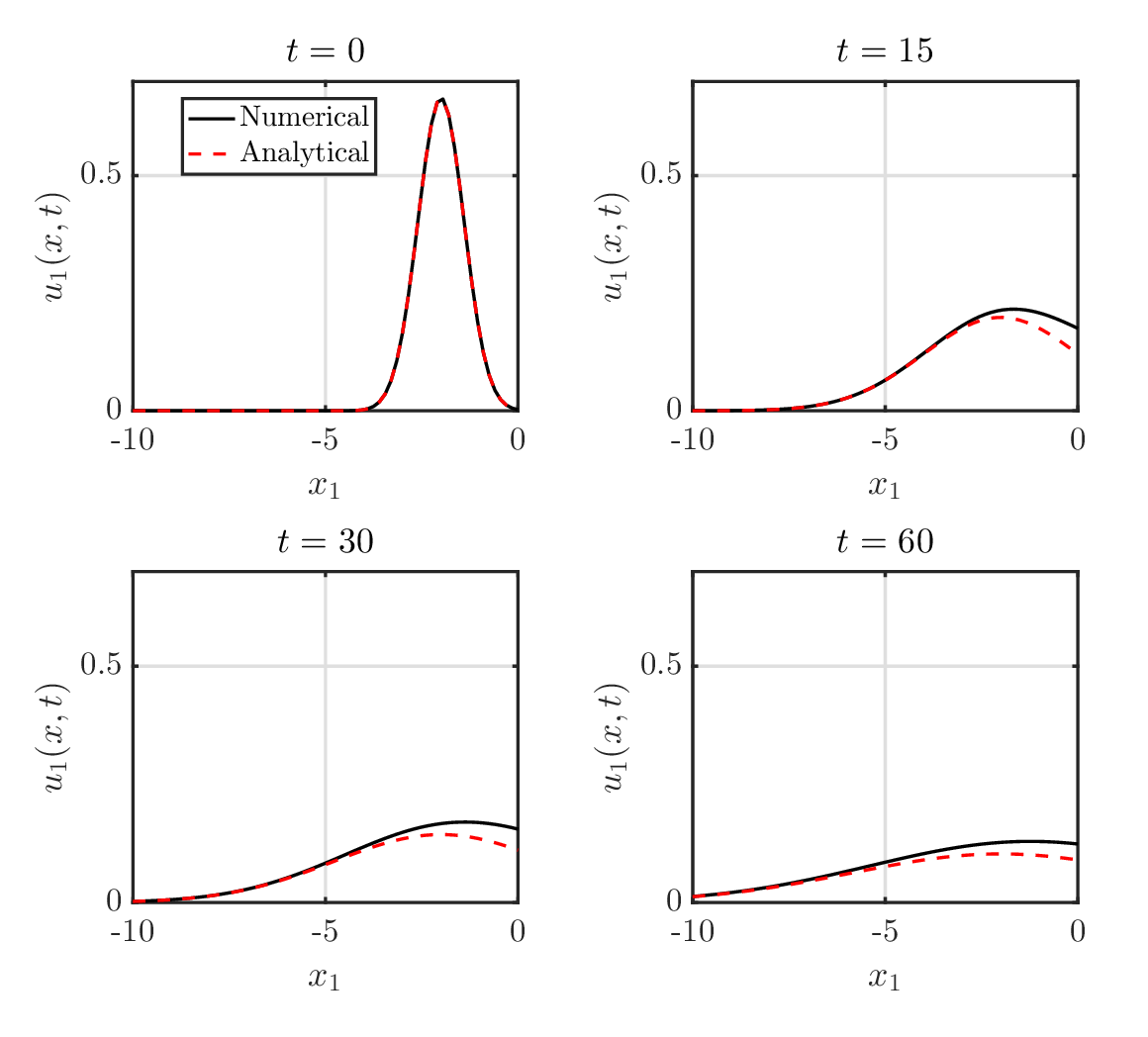}
\caption{
Comparison of the numerical and analytical temperature profiles on bond~1,
$x_1\in[-L,0]$, at $t=0$, $15$, $30$, and $60$, when the sum rule
$\kappa_1=\kappa_2+\kappa_3$ is violated  $\kappa_3=0.35$ 
$\kappa_2=0.089$, and $\kappa_3=0.08$. 
The solid black curves denote the numerical CN
solution, and the dashed red curves correspond to the
analytical Gaussian solution on the real line with $\kappa_1= 0.35$. }
\label{fig:tmprs2}
\end{figure}

\begin{figure}[t]
\centering
\includegraphics[width=1.\textwidth]{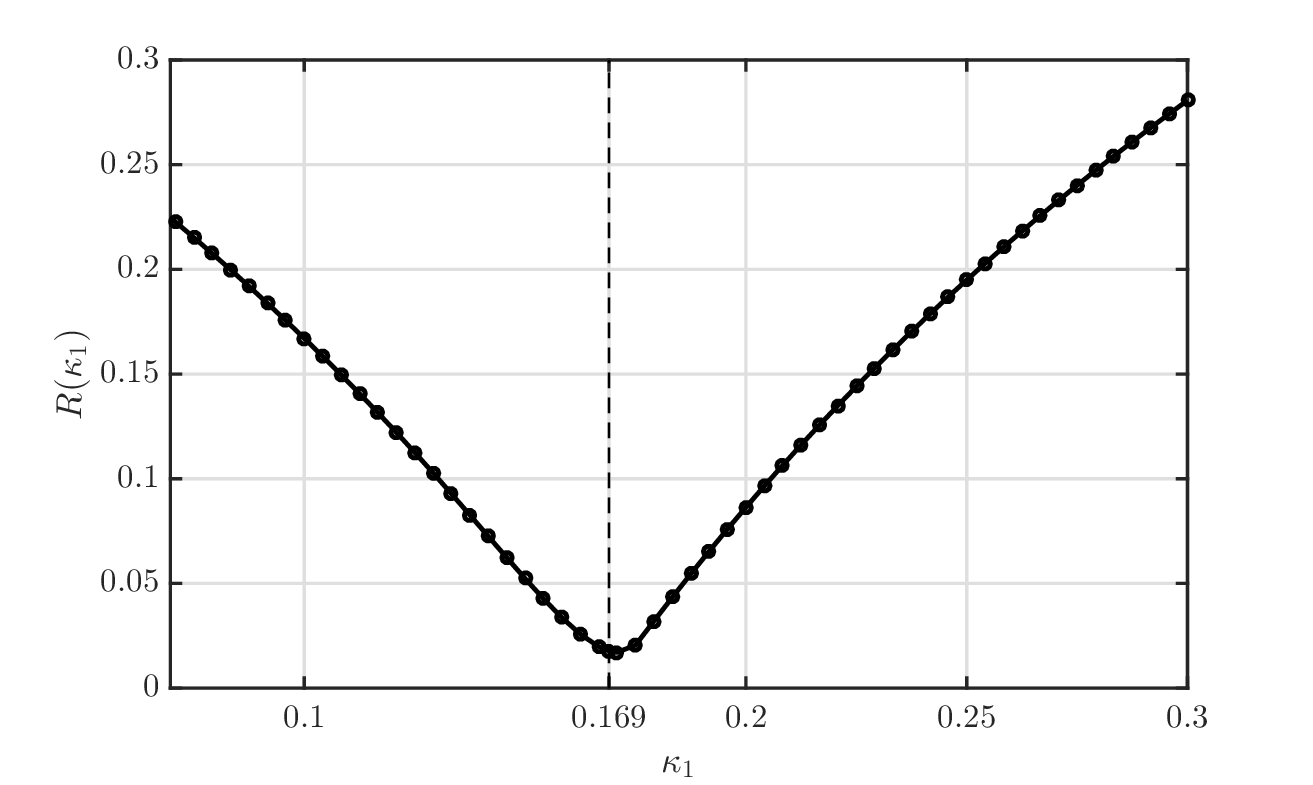}
\caption{
 The backflow rate, $R$ from Eq.~\eqref{rrate}, as a function of $\kappa_1$ for fixed values of $\kappa_2 =0.089$, $\kappa_3 =0.08$.}
\label{fig:refl}
\end{figure}

The temporal evolution of the temperature profile 
is illustrated in Figure~\ref{fig:temp_profile}, under the regime where the algebraic sum rule~\eqref{sum_r}  
is strictly satisfied.
As 
observed in the numerical snapshots, the incoming thermal distribution is split and transmitted through the vertex without any visible backflow or numerical artifacts at the junction. 
This excellent agreement between the analytical prediction and the numerical results confirms that satisfying the sum rule~\eqref{sum_r} effectively establishes an exact transparent junction for thermal networks.



Figure~\ref{fig:temp_profile2} presents plots of the temperature profile when the sum rule in \eqref{sum_r} is violated for heat conductance values $\kappa_1=0.35$, while $\kappa_2=0.089$ and $\kappa_3=0.08$. Unlike the plots in Figure~\ref{fig:temp_profile}, where the temperature profile spreads without obstruction, the plot in Figure~\ref{fig:temp_profile2} exhibits a slight slowing of the spreading process, indicative of a certain backflow. 

Figures~\ref{fig:tmprs} and \ref{fig:tmprs2} compare the numerically computed temperature profiles on the first bond with the corresponding analytical solution on the whole real line, for the cases in which the sum rule is satisfied and violated, respectively. 
When $\kappa_1=\kappa_2+\kappa_3$, the vertex conditions are expected to be transparent to the incoming heat pulse. 
Consequently, the solution on the first bond should coincide with the free-space analytical solution. The excellent agreement observed throughout the evolution confirms this transparency of the vertex. 
In contrast, when the sum rule is violated, the numerical and analytical profiles agree initially, but a noticeable deviation develops as the heat pulse approaches and passes through the vertex. 
The numerical profile exhibits a slight slowing of the spreading, indicating the presence of backflow induced by the junction. 
This deviation demonstrates that, in the absence of the sum rule, the vertex is no longer transparent and affects the heat propagation along the first bond.

To further verify this observation, Figure~\ref{fig:refl} shows the backflow rate, defined by
\begin{equation}\label{rrate}
    R(\kappa_1)=\frac{Q_1}{Q_1+Q_2+Q_3},
\end{equation}
where
\begin{equation*}
    Q_j(t) = \int_0^t q_j(0,\tau)\,d\tau,
\end{equation*}
and the partial heat fluxes $q_j(x_j,t)$ are given by
\begin{equation}
    q_j(x_j,t)=-\kappa_j^2\frac{\partial u_j(x_j,t)}{\partial x_j}.
\end{equation}

\begin{figure}[t]
\centering
\includegraphics[width=1.05\textwidth]{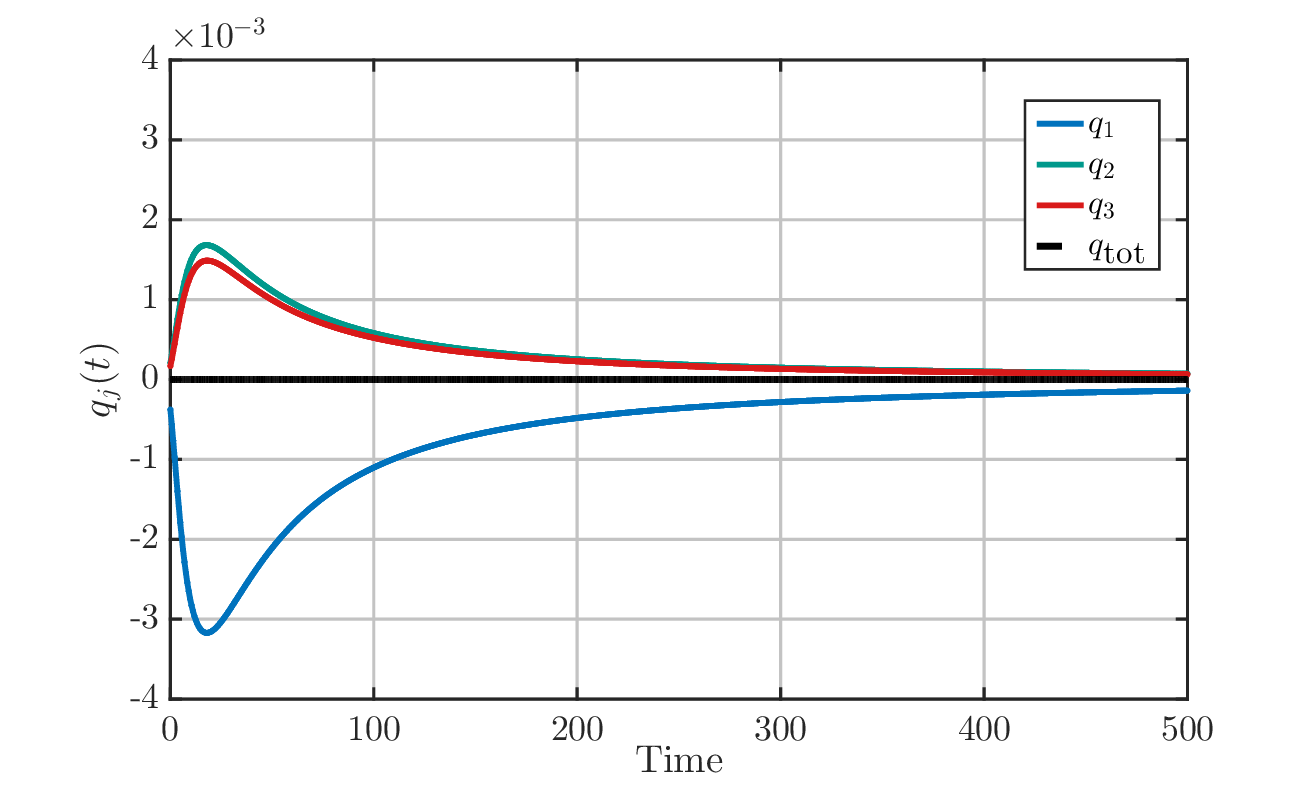}
\caption{Time evolution of the heat flux at the vertex of the star graph in the reflectionless regime corresponding to the sum rule~\eqref{sum_r}, $\kappa_1 = \kappa_2 + \kappa_3=0.169$ with $\kappa_2=0.089$, and $\kappa_3=0.08$. 
The blue, red, and yellow curves represent the heat fluxes on bonds~1, 2, and~3, respectively, while the dashed black curve denotes the total heat flux.}
\label{fig:heat_flux}
\end{figure}

The quantity $R(\kappa_1)$ is plotted as a function of $\kappa_1$, while $\kappa_2$ and $\kappa_3$ are held constant. As can be seen in the figure, $R(\kappa_1)$ reaches its minimum at the value of $\kappa_1$ for which the sum rule in Eq.~\eqref{sum_r} is satisfied. Furthermore, the minimum value is very close to zero, approximately $10^{-3}$, indicating that the backflow is effectively suppressed when the sum rule is satisfied.

Additionally, Figure~\ref{fig:heat_flux} presents the time evolution of the heat fluxes at the graph vertex along each bond when the sum rule is satisfied. The incoming flux along bond 1 is balanced by the outgoing fluxes along bonds 2 and 3. As expected, the total flux at the vertex remains approximately zero throughout the evolution. 
However, this balance does not result from the sum rule. 
Rather, it follows directly from the Kirchhoff flux-conservation boundary condition imposed at the vertex, and therefore holds irrespective of whether the sum rule is satisfied.

\begin{figure}[t]
\centering
\includegraphics[width=1.05\textwidth]{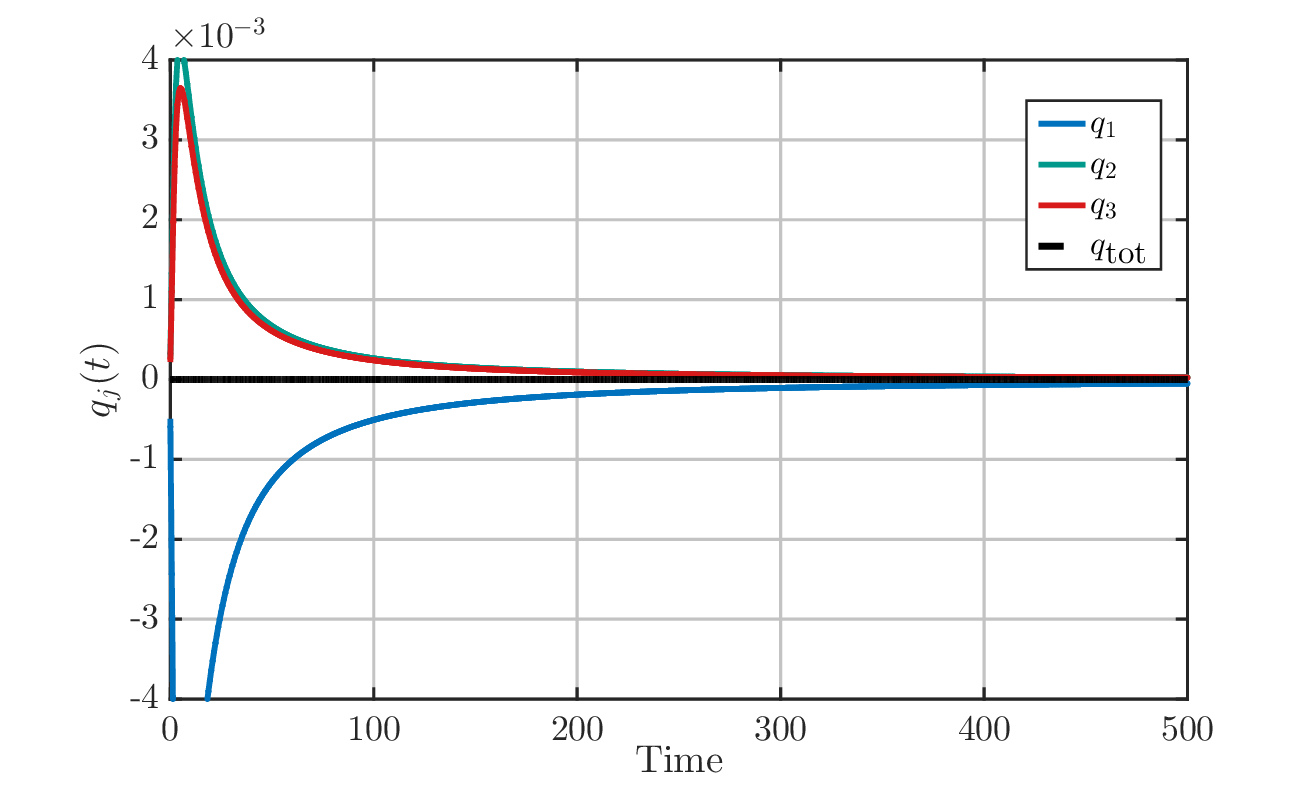}
\caption{Time evolution of the heat flux at the vertex of the star graph when the sum rule~\eqref{sum_r} is broken, $\kappa_1 \ne \kappa_2 + \kappa_3$, with $\kappa_3=0.35$ 
$\kappa_2=0.089$, and $\kappa_3=0.08$. 
The blue, red, and yellow curves represent the heat fluxes on bonds~1, 2, and~3, respectively, whereas the dashed black curve shows the sum of the heat fluxes over all three bonds.}
\label{fig:heat_flux_broken}
\end{figure}

When the sum rule is violated, the corresponding vertex heat fluxes are shown in Figure~\ref{fig:heat_flux_broken}. Again, the total flux remains approximately zero, as required by the imposed Kirchhoff condition. Nevertheless, the individual fluxes are redistributed differently among the three bonds, reflecting the loss of transparency at the vertex. 
Thus, violating the sum rule does not imply a violation of heat-flux conservation. Instead, it alters the transmission properties of the vertex and leads to the non-transparent behavior observed in Figures~\ref{fig:temp_profile2} and \ref{fig:tmprs2}.

\section{Summary and Discussion}\label{sec5}
In this work, we presented a new mathematical model for tunable thermal diffusion, ensuring lack of  thermal backflow across a branching junction within a network structure. 
We successfully simulated the behavior of a time-dependent heat equation on a three-bond star graph topology by combining the principles of TBCs with PDEs defined on metric graphs.
Our core analytical finding is the derivation of an explicit algebraic constraint on the network’s physical properties, expressed as a simple sum rule for diffusion coefficients: 
$\kappa_1 = \kappa_2 + \kappa_3$.
When this parameter relation is satisfied, the non-local convolution-type boundary operator at the junction simplifies to perfectly match an unbounded domain. 
This eliminates thermal backflow and ensures that the heat flux splits seamlessly from the incoming channel into the outgoing branches, generating no artificial reflections at the network nodes.

To validate our analytical model, we implemented an unconditionally stable Crank-Nicolson FDM to solve the heat equation on a network. 
Instead of requiring a complex, explicit discretization of a non-local TBC operator at an external boundary, our approach elegantly restricts discretization to local heat equations on the bonds. 
This approach enforces exact transparent transmission via standard continuity and Kirchhoff-type flux balance rules at the vertex. 
Numerical results confirmed the smooth propagation of an initially localized thermal pulse through the junction and the absence of thermal backflow. 

These findings have several significant implications and open up avenues for future research.
In the fields of nanotechnology and microfluidics, minimizing thermal energy losses and preventing unwanted back diffusion at structural interfaces is essential for optimizing and designing highly efficient thermoelectric devices and nano\-structured carbon materials \cite{cahill, Dhar2008}. 
From a control systems perspective, the derived sum rule offers a simple design constraint for material engineering, allowing for predictable, optimized thermal routing in low-dimensional structures. 
Additionally, although this methodology was demonstrated on a basic star graph, it can be applied to more complex network architectures, such as closed loops, tree-like structures, and multi-junction polymer or circuit models \cite{cahill, Dhar2008}.
This extension builds on existing TBC frameworks for advanced metric graph topologies \cite{Adami2016} and bridges an important gap in modeling diffusion.

\section*{Acknowledgements} 
JM thanks the German Academic Exchange Service (DAAD) for providing a scholarship under the programme “Research Grants in Germany, 2026” (Programme ID No. 57812126), which supported his short-term research stay at the University of Wuppertal.
This work is partially supported by a grant from the Innovation Development Agency of the Republic of Uzbekistan (Ref.\ No.\ F-2021-440).


\end{document}